\documentstyle[12pt,epsfig]{article}
\newcommand{\bea}{\begin{eqnarray}}
\newcommand{\eea}{\end{eqnarray}}
\newcommand{\be}{\begin{equation}}
\newcommand{\ee}{\end{equation}}

\begin{document}

\begin{center}

{\bfseries RHO-LIKE MESONS FROM ANALYSIS OF THE PION-PION SCATTERING}

\vskip 5mm

Yu.S.~Surovtsev$^{a}$ and P.~Byd\v{z}ovsk\'y$^b$

\vskip 5mm
{\small
(a) {\it
Bogoliubov Laboratory of Theoretical Physics, JINR, Dubna 141 980, Russia
}
\\
(b) {\it
Nuclear Physics Institute, ASCR, 25068 \v{R}e\v{z}, Czech Republic
}}
\end{center}

\begin{abstract}
Considering analyticity, unitarity and an influence of
coupled channels, experimental data on the isovector $P$-wave of $\pi\pi$
scattering was analyzed to study $\rho$-like mesons below 1900 MeV.
The analysis indicates evidently that in the energy range 1200--1800 MeV, 
there are three $\rho$-like mesons: $\rho(1250)$, $\rho(1450)$ and 
$\rho(1600)$, unlike the PDG tables. The obtained P-wave $\pi\pi$-scattering 
length ($a_1^1=33.9\pm2.02~[10^{-3}m_{\pi^+}^{-3}]$) most matches to the one 
calculated in the local Nambu--Jona-Lasinio model.
\end{abstract}

\section{Introduction}

The investigation of vector mesons is actual up to now due to their role 
in forming the electromagnetic structure of particles and because, {\it e.g.}, 
in the $\rho$-family, only the $\rho(770)$ meson can be deemed to be well 
understood \cite{PDG-06}. The other $\rho$-like mesons must be either still 
confirmed in various experiments and analyses or their parameters essentially 
corrected. For example, the $\rho(1250)$ meson which was discussed actively 
some time ago \cite{BBS,GG} was confirmed relatively recently in the amplitude 
analysis of the LASS Collaboration \cite{Aston-LASS} and in combined analysis 
of several processes \cite{Henner}. However this state is referred to only 
slightly in the PDG issue \cite{PDG-06} (the relevant observations are listed 
under the $\rho(1450)$).

On the other hand, the $\pi\pi$ interaction plays a central role in physics of
strongly interacting particles and, therefore, it has always been an object of
continuous investigation. Let us note only some recent works devoted also to the
theoretical study of the isovector $P$-wave of $\pi\pi$ scattering. First, there 
are the analyses of available experimental data on the $\pi\pi$ scattering 
utilizing the Roy equations \cite{Leutwyler,KLL,CCL} and the forward dispersion 
relations \cite{Yndurain,Kaminski}, in which, {\it e.g.},  low-energy
parameters of the $\pi\pi$ scattering were obtained. Second, there are the works 
in which the low-energy parameters are calculated in chiral theories with 
the linear realization of chiral symmetry \cite{Volkov,BOM}.

We use our model-independent method \cite{KMS-nc96}. It is based on the first
principles (analyticity and unitarity) directly applied to analysis of experimental
data. Our aim is to study the $\rho$-like mesons below 1900 MeV and to obtain the
$\pi\pi$-scattering length. Unfortunately this method, using essentially an
uniformizing variable, is applicable only to the 2-channel case and under some
conditions to the 3-channel one.
Here the thresholds of the $\pi\pi$ and $\omega\pi$ channels are allowed for
explicitly in the uniformizing variable (in the threshold region of the second
channel, one has observed a deviation from elasticity of the $P$-wave $\pi\pi$
scattering). Influence of other coupled channels is supposed to be taken into
account through the background. In order to investigate a coupling of
resonances with these other channels, we also apply multichannel Breit--Wigner
forms to generate the resonance poles.

The paper is organized as follows. In Section II, we outline the method of the
uniformizing variable in applying it to studying the 2-channel $\pi\pi$
scattering and present results of the analysis of the available data
\cite{Protopopescu}--\cite{Estabrooks} on the isovector $P$-wave of $\pi\pi$
scattering. Section III is devoted to analysis of the same data using the
Breit--Wigner forms. Finally, in Section VI, we summarize and discuss obtained
results.

\section{Analysis of $P$-wave $\pi\pi$ scattering in the uniformizing 
variable method}

Let the $\pi\pi$-scattering $S$-matrix be determined on the 4-sheeted Riemann
surface with the right-hand branch-points at $4m_{\pi^+}^2$ and
$(m_\omega+m_{\pi^0})^2$ and also with the left-hand one at $s=0$. It is supposed
that influence of other branch points can be taken into account through the
background. The Riemann-surface sheets are numbered according to the signs of
analytic continuations of the channel momenta
$$k_1=\frac{1}{2}\sqrt{s-4m_{\pi^+}^2} ~~~{\rm and}~~~
k_2=\frac{1}{2}\sqrt{s-(m_\omega+m_{\pi^0})^2}$$ as follows:
$\mbox{signs}({\mbox{Im}}k_1,{\mbox{Im}}k_2)= ++,-+,--$ and $+-$~
correspond to sheets I, II, III and IV, respectively.

The $S$-matrix is supposed to be ~$S=S_{res}S_{bg}$ where $S_{res}$ represents
resonances and $S_{bg}$, the background. In general, an explicit allowance for
the $(m_\omega+m_{\pi^0})^2$ branch point would permit us to describe transitions
between the $\pi\pi$ and $\omega\pi$ initial and final states with the help of
the only one function $d(k_1,k_2)$ ~(the Jost matrix determinant) using the
Le Couteur--Newton relations \cite{LN}. Unfortunately, the data on process
$\pi\pi\to\omega\pi$ are absent.

In Ref. \cite{KMS-nc96} it was shown how one can obtain the multichannel
resonance representations by poles and zeros on the Riemann surface with the 
help of the formulas, expressing analytic continuations of the matrix elements,
describing the coupled processes, to unphysical sheets in terms of those on sheet
I. It is convenient to start from resonance zeros on sheet I. Then in the
2-channel $\pi\pi$ scattering, we have {\it three types} of resonances:
({\bf a}) described by a pair of complex conjugate zeros in the $S$-matrix
element on sheet I and by a pair of conjugate shifted zeros on sheet IV;
({\bf b}) described by a pair of conjugate zeros on sheet III and by a pair of
conjugate shifted zeros on sheet IV;
({\bf c}) which correspond to a pair of conjugate zeros on sheet I, the one on
sheet III and two pairs of conjugate zeros on sheet IV. Due to unitarity the
poles on sheet II, III and IV are situated in the same energy points as the
corresponding zeros on sheet I, IV and III, respectively. Note that the size of
shift of zeros on sheet IV relative to the ones on sheets I and III is determined
by the strength of coupling of the channels (here $\pi\pi$ and $\omega\pi$). The
cluster kind is related to the nature of resonance.

With the help of the uniformizing variable\footnote{The analogous uniformizing
variable has been used, {\it e.g.}, in Ref. \cite{Meshch} in studying the forward
elastic $p{\bar p}$ scattering amplitude and in Ref. \cite{SKN-epja02} 
in the combined analysis of data on processes $\pi\pi\to\pi\pi,K\overline{K}$ 
in the channel with $I^GJ^{PC}=0^+0^{++}$.}
\begin{equation} 
\label{v}
v=\frac{(m_\omega+m_{\pi^0})/2~\sqrt{s-4m_{\pi^+}^2} +
m_{\pi^+}~\sqrt{s-(m_\omega+m_{\pi^0})^2}}{\sqrt{s\left[\left((m_\omega+
m_{\pi^0})/2\right)^2-m_{\pi^+}^2\right]}},
\end{equation}
the considered 4-sheeted Riemann surface is mapped onto the $v$-plane, divided
into two parts by a unit circle centered at the origin. Sheets I (II), III (IV)
are mapped onto the exterior (interior) of the unit disk on the upper and lower
$v$-half-plane, respectively. The physical region extends from the point $i$ on
the imaginary axis ($\pi\pi$ threshold) along the unit circle clockwise in the
1st quadrant to the point 1 on the real axis ($\omega\pi^0$ threshold) and then
along the real axis to the point
$b=\sqrt{(m_\omega+m_{\pi^0}+2m_{\pi^+})/(m_\omega+m_{\pi^0}-2m_{\pi^+})}$ 
into which $s=\infty$ is mapped on the $v$-plane. The intervals
$(-\infty,-b],[-b^{-1},b^{-1}],[b,\infty)$ on the real axis are the images of the
corresponding edges of the left-hand cut of the $\pi\pi$-scattering amplitude.
The ({\bf a}) resonance is represented in $S(\pi\pi\to\pi\pi)$ by two pairs of
poles on the images of sheets II and III, symmetric to each other with respect to
the imaginary axis, and by zeros, symmetric to these poles with respect to the
unit circle. Note that a symmetry of the zeros and poles with respect to the
imaginary axis appears due to the real analyticity of the $S$-matrix, and the
symmetry of the poles and zeros with respect to the unit circle ensures a
realization of the known experimental fact that the $\pi\pi$ interaction is
practically elastic up to a vicinity of the $\omega\pi^0$ threshold.

The resonance part of $S$-matrix $S_{res}$ becomes {\it a one-valued function}
on the $v$-plane and, in the $\pi\pi$ channel, it is expressed through the 
$d(v)$-function as follows\footnote{Other authors also have used 
the parameterizations with the Jost functions at analyzing the $S$-wave 
$\pi\pi$ scattering in the one-channel approach \cite{Bohacik} and in the 
two-channel one \cite{MP-93}. In latter work, the uniformizing variable $k_2$ 
has been used and the $\pi\pi$-threshold branch-point has been neglected, 
therefore, their approach cannot be employed near by the $\pi\pi$ threshold.}:
\begin{equation} \label{v:S}
S_{res}=\frac{d(-v^{-1})}{d(v)}
\end{equation}
where $d(v)$ represents the contribution of resonances, described by one of three
types of the pole clusters in the 2-channel case, {\it i.e.},
\begin{equation} \label{d_res}
d = v^{-M}\prod_{n=1}^{M} (1-v_n^* v)(1+v_n v)
\end{equation}
with $M$ the number of pairs of the conjugate zeros.

The background part $S_{bg}$ is taken in the form:
\begin{equation} \label{S_bg}
S_{bg}=\exp\left[2i\left(\sqrt{\frac{s-4m_{\pi^+}^2}{s}}\right)^{3}\left(\alpha_1
+ \alpha_2~\frac{s-s_1}{s}~\theta(s-s_1)
+ \alpha_3~\frac{s-s_2}{s}~\theta(s-s_2)\right)\right]
\end{equation}
where $\alpha_i=a_i+ib_i$, $s_1$ is the threshold of 4$\pi$ channel noticeable in
the $\rho$-like meson decays, $s_2$ is the threshold of $\rho2\pi$ channel. Due
to allowing for the left-hand branch-point at $s=0$ in the uniformizing variable
(\ref{v}), $a_1=b_1=0$. Furthermore, $b_2=0$ is an experimental fact.

With formulas (\ref{v:S})--(\ref{S_bg}), we have analyzed data
\cite{Protopopescu}--\cite{Estabrooks} for the inelasticity parameter ($\eta$)
and phase shift of the $\pi\pi$-scattering amplitude ($\delta$)
($S(\pi\pi\to\pi\pi)=\eta\exp(2i\delta)$), introducing three ($\rho(770)$,
$\rho(1250-1580)$ and $\rho(1550-1780)$), four (the indicated ones plus
$\rho(1860-1910)$) and five (the indicated four plus $\rho(1450)$) resonances.
In Ref.\cite{Estabrooks}, results of two analyses are cited: one uses the
$s$-channel helicity amplitude when extracting the $\pi\pi$-scattering amplitude
on the $\pi$-exchange pole; in the other, the $t$-channel one is used instead.
Therefore, we have taken both analyses as independent. There are given the data
for the phase shift of amplitude below the $K\overline{K}$ threshold. Comparing
these data with the ones of Refs.\cite{Protopopescu,Hyams}, one can see that the
points of the former lie systematically by 1$^\circ$-5$^\circ$ higher than the
ones of the latter, except for two points of Ref.\cite{Protopopescu} at 710 and
730~MeV, which lie by about 2$^\circ$ higher than the corresponding points of
Ref.\cite{Estabrooks} and which are omitted in the subsequent analyses. Since we
do not know the energy dependence of the remarked deviations of points, we have
supposed a constant systematic error that must be determined in the combined
analysis of data. We have obtained a satisfactory description with
$\chi^2/\mbox{NDF}$ and with the indicated systematic error equal respectively to
$293.518/(186-15)=1.716$ and $-1.885^\circ$ for the case of three resonances,
$280.043/(186-19)=1.677$ and $-1.897^\circ$ for four resonances,
$269.574/(186-23)=1.654$ and $-1.876^\circ$ for five resonances. When calculating
$\chi^2$ for the inelasticity parameter, three points of data \cite{Hyams} at 990,
1506 and 1825 MeV have been omitted in all three cases as giving the anomalously
big contribution to $\chi^2$. When calculating $\chi^2$ for the phase shift,
three points of data \cite{Estabrooks} have been omitted in all three cases: the
one at 790 MeV from the $s$-channel analysis, and two at 790 and 850 MeV from the
$t$-channel one.

\begin{figure}[h]
\begin{center}
%\hspace*{-0.9cm}
\epsfig{file=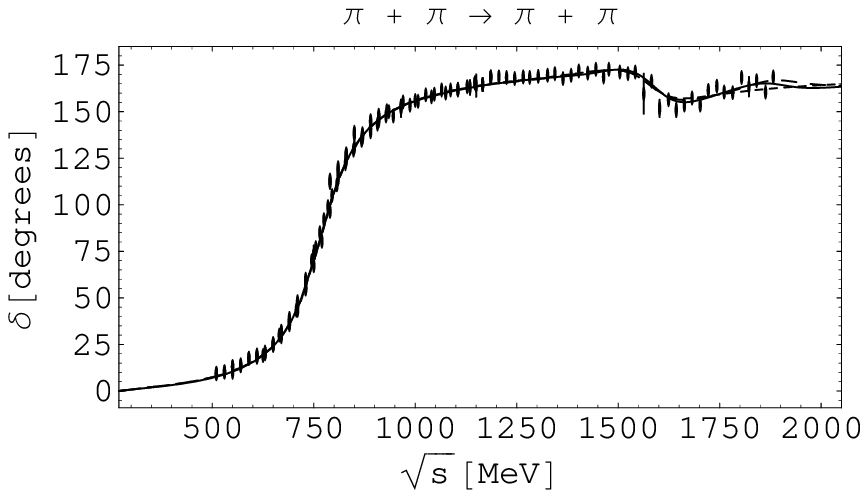,width=11cm}
%\hspace*{-1.1cm}
\epsfig{file=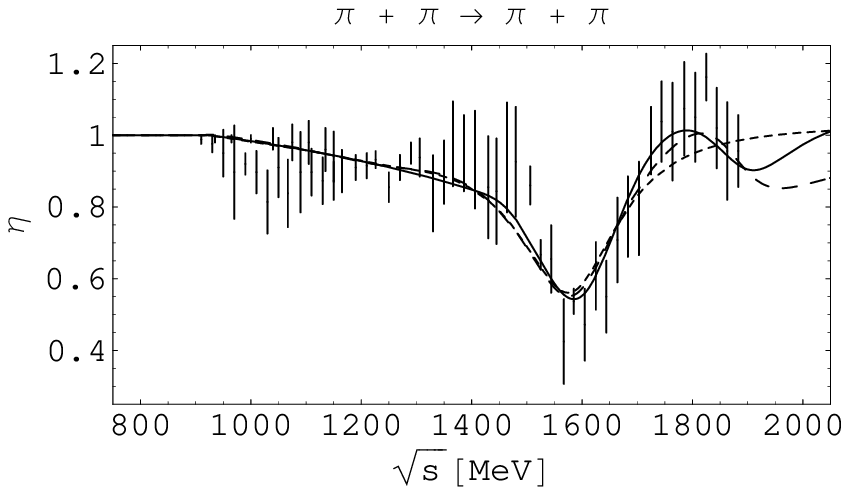,width=11cm}
\end{center}
\vskip -.3cm
\caption{The phase shift of amplitude (the upper figure) and module of matrix 
element (the lower figure) of the $P$-wave $\pi\pi$-scattering. The results 
of 3-resonance (short-dashed), 4-resonance (long-dashed), and 5-resonance (solid) 
descriptions are shown. The data from Refs. \cite{Protopopescu}--\cite{Estabrooks}.}
\label{fig:m-ind.phs.mdl}
\end{figure}

On figures \ref{fig:m-ind.phs.mdl}, we demonstrate results from our fitting to
data \cite{Protopopescu}--\cite{Estabrooks}. The short-dashed curves are for the
three-resonance description, the long-dashed ones for the four-resonance and the
solid for the five-resonance cases.

From a variety of possible resonance representations by pole-clusters, the
analyses prefer he following one to be the most relevant: {\it the $\rho(770)$ is
described by the cluster of type ({\bf a}) and the others by type ({\bf b})}.

Let us show the obtained pole clusters on the lower $\sqrt{s}$-half-plane in
Table~\ref{clusters} (it is clear there are also complex-conjugate poles on 
the upper half-plane).
\begin{table}[h] \centering\caption{Pole clusters for the $\rho$-like resonances.}
\vskip0.3truecm
\def\arraystretch{1.5}
\begin{tabular}{|c|c|c|c|c|}\hline\multicolumn{5}{|c|} {Three resonances}\\
\hline \multicolumn{2}{|c|}{Sheet} &
II & III & IV \\ \hline {$\rho(770)$} & {$\sqrt{s_r}$, MeV} &
$767.3-i73.3$ & $782-i65.6$ & {} \\
\hline {$\rho(1250)$} & {$\sqrt{s_r}$, MeV} &
{} & $1249.9-i152$ & $1249-i146.2$ \\
\hline {$\rho(1600)$} & {$\sqrt{s_r}$, MeV}
& {} & $1585-i130.5$ & $1578-i72.2$ \\
\hline
\multicolumn{5}{|c|} {Four resonances}\\
\hline \multicolumn{2}{|c|}{Sheet} &
II & III & IV \\ \hline {$\rho(770)$} & {$\sqrt{s_r}$, MeV} &
$766.5-i73.2$ & $783.1-i66.2$ & {} \\
\hline {$\rho(1250)$} & {$\sqrt{s_r}$, MeV} &
{} & $1251.4-i152.1$ & $1249-i144.3$ \\
\hline {$\rho(1600)$} & {$\sqrt{s_r}$, MeV}
& {} & $1585.2-i141.8$ & $1579.6-i73.6$ \\
\hline {$\rho(1900)$} & {$\sqrt{s_r}$, MeV} &
{} & $1871.5-i97.2$ & $1894-i95.3$ \\
\hline
\multicolumn{5}{|c|} {Five resonances}\\
\hline \multicolumn{2}{|c|}{Sheet} &
II & III & IV \\ \hline {$\rho(770)$} & {$\sqrt{s_r}$, MeV} &
$765.8-i73.3$ & $778.2-i68.9$ & {} \\
\hline {$\rho(1250)$} & {$\sqrt{s_r}$, MeV} &
{} & $1250-i131.4$ & $1249.4-i130.7$ \\
\hline {$\rho(1470)$} & {$\sqrt{s_r}$, MeV} &
{} & $1469.2-i89.3$ & $1465.4-i100.4$ \\
\hline {$\rho(1600)$} & {$\sqrt{s_r}$, MeV}
& {} & $1634.8-i145.9$ & $1593.4-i72.9$ \\
\hline {$\rho(1900)$} & {$\sqrt{s_r}$, MeV} &
{} & $1883-i106.5$ & $1893.4-i87.6$ \\
\hline
\end{tabular}
\label{clusters}
\end{table}
The background parameters are: $a_2=0.0093$, $a_3=0.0618$ and $b_3=-0.0135$ for
the three-resonance, $a_2=0.00166$, $a_3=0.0433$ and $b_3=-0.00442$ for the
four-resonance and $a_2=0.0248$, $a_3=0.0841$ and $b_3=0.0019$ for the
five-resonance descriptions.

Though the description can be considered practically as the same in all three
cases, careful consideration of the obtained parameters and energy dependence of
the fitted quantities suggests that {\it the resonance $\rho(1900)$ strongly
desired and that the $\rho(1450)$ should be also present improving slightly the
description}.

Masses and widths of the obtained $\rho$-states can be calculated from the pole
positions on sheet II for resonances of type ({\bf a}) and on sheet IV for 
resonances of type ({\bf b}). If the resonance part of the amplitude reads as
$$T^{res}=\frac{\sqrt{s}~\Gamma_{el}}{m_{res}^2-s-i\sqrt{s}~\Gamma_{tot}},$$
we obtain for the masses and total widths, respectively, the following
values (in the MeV units):
\begin{center}
{\begin{tabular}{lllll} for & ~~$\rho(770)$, & ~~769.3 & ~~and & ~~146.6;\\
for & ~~$\rho(1250)$, & ~~1256.2 & ~~and & ~~261.4;\\
for & ~~$\rho(1470)$, & ~~1468.8 & ~~and & ~~200.8;\\
for & ~~$\rho(1600)$, & ~~1595.1 & ~~and & ~~145.8;\\
for & ~~$\rho(1900)$, & ~~1895.4 & ~~and & ~~175.2.\end{tabular}}\end{center}

\section{The Breit--Wigner analysis of $P$-wave $\pi\pi$ scattering}

In various works \cite{PDG-06}, it was shown that the $\rho$-like resonances
obtained in the previous section have also other considerable decay channels 
in addition to those considered explicitly above. It was observed that the
$\rho(1450)$ and/or a possible $\rho(1250)$ can also decay to the $\eta\rho^0$ 
($< 4\%$), $\phi\pi$ ($< 1\%$), and $4\pi$ (seen) channels, where the fraction 
$\Gamma_i/\Gamma$ \cite{PDG-06} is given in parenthesis. The $\rho(1700)$ 
resonance has the large branching ratio to the $4\pi$ (large), $\rho2\pi$ 
(dominant) and $\eta\rho^0$ (seen) channels.

To include explicitly influence of some selected channels and to obtain
information about couplings with these channels on the basis of analysis of 
the $\pi\pi$-scattering data, we have used 5-channel Breit--Wigner forms in
constructing the Jost matrix determinant $d(k_1,\cdots,k_5)$. To generate the
resonance poles and zeros in the $S$-matrix, the Le~Couteur--Newton relation
was utilized:
\begin{equation} \label{CN:S}
S_{res}=\frac{d(-k_1,\cdots,k_5)}{d(k_1,\cdots,k_5)}.
\end{equation}
In eq.(\ref{CN:S}) $k_1$, $k_2$, $k_3$, $k_4$ and $k_5$ are the $\pi\pi$-,
$\pi^+\pi^-2\pi^0$-, $2(\pi^+\pi^-)$-, $\eta2\pi$- and $\omega\pi^0$-channel
momenta, respectively. The $d$-function is taken as ~$d=d_{res}d_{bg}$ with
$d_{res}$, describing resonance contributions, and $d_{bg}$, the background.

The Breit--Wigner form for the resonance part of the $d$-function is assumed as
\begin{equation}
d_{res}(s)=\prod_{r}
\left[M_r^2-s-i\sum_{j=1}^5\rho_{rj}^3~R_{rj}~f_{rj}^2\right],
\end{equation}
where $\rho_{rj}=k_i(s)/k_i(M_r^2)$, $f_{rj}^2/M_r$ is the partial width 
of resonance with mass $M_r$, $R_{rj}$ is a Blatt--Weisskopf barrier 
factor \cite{Blatt-Weisskopf} conditioned by the resonance spins. 
For the vector particle this factor have the form:
\begin{equation}
R_{rj}=\frac{1+\frac{1}{4}(\sqrt{M_r^2-4m_j^2}~r_{rj})^2}
{1+\frac{1}{4}(\sqrt{s-4m_j^2}~r_{rj})^2}
\end{equation}
with radius $r_{rj}=0.7035$ fm which in our analysis are equal for all 
resonances in all channels. Furthermore, we have assumed that the widths 
of resonance decays to the $\pi^+\pi^-2\pi^0$ and $2(\pi^+\pi^-)$ channels 
are related by relation: $f_{r2}=f_{r3}/\sqrt{2}$.

The background part $d_{bg}$ is
\begin{equation} \label{d bg}
d_{bg}=\exp\left[-i\left(\sqrt{\frac{s-4m_{\pi^0}^2}{s}}\right)^3\left(\alpha_1+
\alpha_2~\frac{s-s_1}{s}~\theta(s-s_1)\right)
\right]
\end{equation}
where $\alpha_i=a_i+ib_i$, $s_1$ is the threshold of $\rho2\pi$ channel (it is
clear that $a_2$ and $b_2$ take into account also influence of other channels
opened at higher energies than the $\rho2\pi$ threshold); $b_1$ is taken to 
be zero.

\begin{figure}[h]
\begin{center}
%\hspace*{-0.9cm}
\epsfig{file=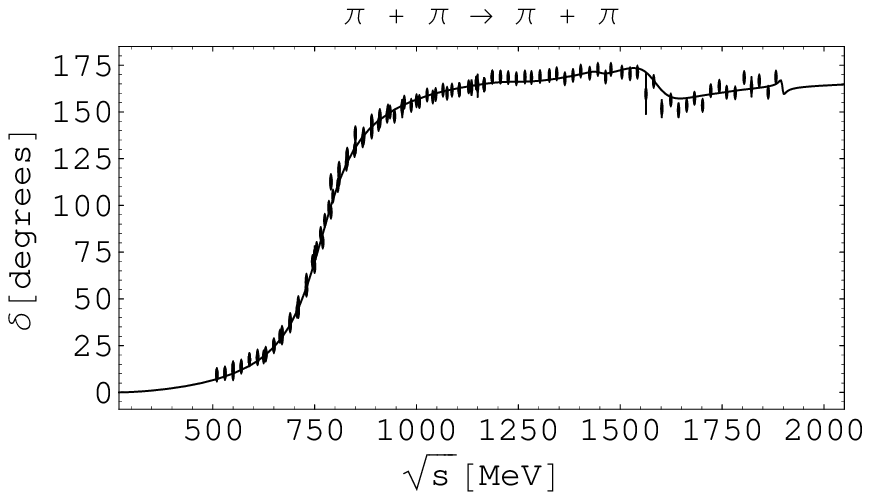,width=11cm}
%\hspace*{-1.1cm}
\epsfig{file=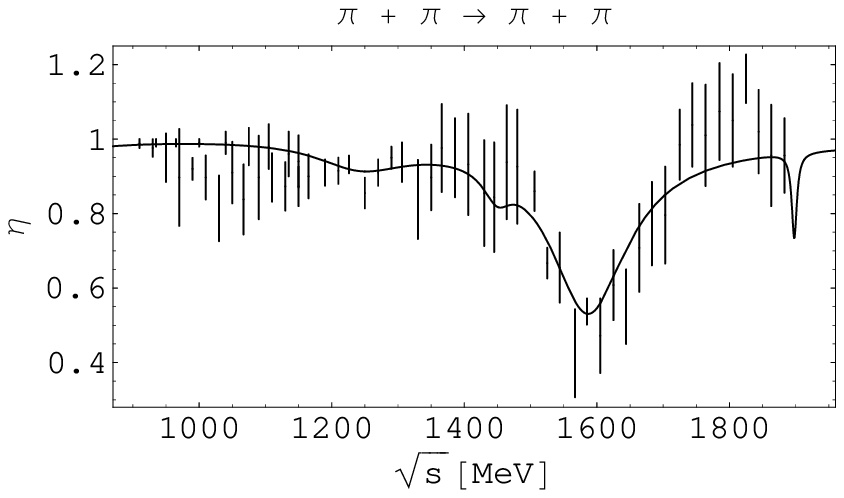,width=11cm}
\end{center}
\vskip -.3cm
\caption{The phase shift of amplitude and module of matrix element of the 
$P$-wave $\pi\pi$-scattering. The curves show result of fitting to data
\cite{Protopopescu}--\cite{Estabrooks} using the Breit-Wigner form.}
\label{fig:BWphs.mdl}
\end{figure}

Using formulas (\ref{CN:S})--(\ref{d bg}) we have performed the analysis, just 
as in the previous section, with three, four and five resonances. We have 
obtained the same reasonable description in all three cases: the total
$\chi^2/\mbox{NDF}=316.206/(186-17)=1.871$, $314.688/(186-22)=1.919$, and
$303.101/(186-27)=1.906$ for the case of three, four, and five resonances, 
respectively. On the figure~\ref{fig:BWphs.mdl}, we demonstrate results of 
our fitting to data \cite{Protopopescu}-\cite{Estabrooks} for the case of five 
resonances.
The systematic error of data \cite{Estabrooks}, discussed in the previous section,
is equal to $-1.987^\circ$ in this case. When calculating $\chi^2$ for the
inelasticity parameter, three points of data \cite{Hyams} at 990, 1030 and 1825
MeV were omitted as giving the anomalously large contribution to $\chi^2$.
When calculating $\chi^2$ for the phase shift, the same three points of data have
been omitted as in the model-independent analysis.
For the background we find: $a_1=-0.00121\pm0.0018$, $a_2=-0.1005\pm0.011$ and
$b_2=0.0012\pm0.006$.

The obtained values of resonance parameters for the last case are given 
in Table~\ref{vparameters}.
\begin{table}[h]
\centering
\caption{The $\rho$-like resonance parameters (all in MeV).}
\vskip0.2truecm
\def\arraystretch{1.5}
\begin{tabular}{|c|c|c|c|c|c|} \hline
{State} &{$\rho(770)$}&{$\rho(1250)$}&{$\rho(1450)$}&{$\rho(1600)$}&{$\rho(1900)$}\\
\hline\hline
~$M$~&777.69$\pm$0.32&1249.8$\pm$15.6&1449.9$\pm$12.2&1587.3$\pm$4.5&1897.8$\pm$38\\
\hline
$f_{r1}$&343.8$\pm$0.73&87.7$\pm$7.4&56.9$\pm$5.4&248.2$\pm$5.2&47.3$\pm$12\\
\hline
$f_{r2}$&24.6$\pm$5.8&186.3$\pm$39.9&100.1$\pm$18.7&240.2$\pm$8.6&73.7\\
\hline
$f_{r3}$&34.8$\pm$8.2&263.5$\pm$56.5&141.6$\pm$26.5&339.7$\pm$12.5&104.3\\
\hline
$f_{r4}$&{}&231.8$\pm$111&141.2$\pm$98&141.8$\pm$33& 9 \\
\hline
$f_{r5}$&{}&231$\pm$115&150$\pm$95&108.6$\pm$40.4& 10 \\
\hline
$\Gamma_{tot}$&$\approx$154.3&$>$175&$>$52&$>$168&$>$10\\
\hline  \end{tabular}
\label{vparameters}
\end{table}
Note that in Table~\ref{vparameters} we do not show the errors for $f_{42}$, 
$f_{43}$, $f_{44}$ and $f_{45}$, because the present data are not sufficient 
to fix reliably these parameters. We may conclude that, using the accessible data,
{\it the model-independent analysis testifies in favour of existence of 
the $\rho(1900)$ and, maybe, $\rho(1450)$, but the Breit--Wigner approach 
cannot verify this result}.

We have also calculated the isovector $P$-wave length of $\pi\pi$ scattering
$a_1^1$. Its value is shown in Table~\ref{pipi.length} in comparison with the
results from various evaluations in the local \cite{BOM} and non-local
\cite{Volkov} Nambu--Jona-Lasinio (NJL) model and from the ones with the use of
Roy's equations \cite{Yndurain,CCL,KLL} (in Ref.\cite{CCL} one apply also the
chiral perturbation theory (ChPT) to construct a precise $\pi\pi$-scattering
amplitude at $s^{1/2}\leq 0.8$ GeV).
\begin{table}[htb] \centering 
\caption{ Comparison of values of the $\pi\pi$ scattering length $a_1^1$ 
from various approaches.}
\vskip0.3truecm
\def\arraystretch{1.5}
\begin{tabular}{|l|c|l|} \hline $a_1^1[10^{-3}m_{\pi^+}^{-3}]$ &
{References} & ~~~~~~~~~~Remarks \\ \hline
\hline $33.9\pm 2.02$ & This paper & Breit--Wigner analysis\\
\hline $ 34 $ & \cite{BOM} & Local NJL model \\
\hline
$ 37 $ & \cite{Volkov} & Non-local NJL model  \\
\hline
$37.9\pm 0.5$ & \cite{CCL} & Roy equations using ChPT \\
\hline
$38.4\pm 0.8$ & \cite{Yndurain} & Roy equations \\
\hline
$39.6\pm 2.4$ & \cite{KLL} & Roy equations \\ \hline
\end{tabular} \label{pipi.length} \end{table}

%\vspace*{-1.5cm}
\section{Conclusions}

The reasonable description of all the accessible experimental data on the
isovector $P$-wave of $\pi\pi$ scattering for the inelasticity parameter ($\eta$)
and phase shift of amplitude ($\delta$) \cite{Protopopescu}-\cite{Estabrooks}
have been obtained up to 1880 MeV based on the first principles (analyticity and
unitarity) directly applied to analysis of the data. Analysis has been carried
out in the model-independent approach using the uniformizing variable (here the
satisfactory description is obtained: $\chi^2/\mbox{NDF}=1.654$) and applying
multichannel Breit--Wigner forms to generate the resonance poles and zeros in the
$S$-matrix ($\chi^2/\mbox{NDF}=1.906$).
The aim of analysis (except for obtaining an unified formula for the $P$-wave
$\pi\pi$ scattering amplitude in the whole of investigated energy range) was to
study  the $\rho$-like mesons below 1900 MeV  and to obtain the $P$-wave
$\pi\pi$-scattering length.

For the $\rho(770)$, the obtained value for mass is a little smaller in the
model-independent approach ($769.3$ MeV) and a little bigger in the Breit--Wigner
one ($777.69\pm0.32$ MeV) than the averaged mass ($775.5\pm0.4$ MeV) cited in the
PDG tables \cite{PDG-06}. However, this mass also occurs in analysis of some 
reactions \cite{PDG-06}.
The obtained value of the total width in the first analysis ($146.6$ MeV) coincides
with the averaged PDG one ($146.4\pm1.1$ MeV) and it is a little bit bigger in
the second analysis ($\approx154.3$ MeV) than the PDG value, however, this
is also encountered in other analyses \cite{PDG-06}. Note that predicted widths
of the $\rho(770)$ decays to the $4\pi$-modes are significantly larger than,
{\it e.g.}, those evaluated in the chiral model of some mesons based on the
hidden local symmetry added with the anomalous terms \cite{Achasov}.

{\it The 2nd $\rho$-like meson has the mass 1256.2 MeV in the 1st analysis and
1249.8$\pm$15.6 MeV in the 2nd one}. This differs significantly from the mass
($1459\pm11$ MeV) of the 2nd $\rho$-like meson cited in the PDG 
tables \cite{PDG-06}. We mentioned already in Introduction that 
the $\rho(1250)$ meson was discussed actively some time ago \cite{BBS,GG}, and 
next the evidence for it was obtained in some analyses \cite{Aston-LASS,Henner}. 
Note also the talk by Ichiro Yamauchi at the HADRON'07 conference \cite{Yamauchi} 
in which this mass value of the second $\rho$-like meson is also confirmed 
in the re-analysis of some data on the $e^+e^-\to\omega\pi^0$ reaction.
To the point, if this state is interpreted as a first radial excitation of the
$1^+1^{--}$ state, then it lies down well on the corresponding linear trajectory
with an universal slope on the $(n,M^2)$ plane \cite{Anisovich} (n is the radial
quantum number of the $q{\bar q}$ state), while the meson with mass $M=1450$ MeV
turns out to be considerably higher than this trajectory.

Further we note that some time ago it was shown that the 1600-MeV region 
contains in fact two $\rho$-like mesons. This conclusion was made on the basis 
of investigation of the consistency of the 2$\pi$ and 4$\pi$ electromagnetic 
form factors and the $\pi\pi$-scattering length \cite{Erkal} and as a result 
of combined analysis of data on the 2$\pi$ and 4$\pi$ final states in $e^+e^-$ 
annihilation and photoproduction \cite{Donnachie87}. We assume this possibility, 
{\it i.e.}, that {\it in the energy range 1200--1800 MeV there are three 
$\rho$-like mesons and the next meson has the mass about 1450 MeV}. This does 
not contradict to data; in the model-independent analysis, description is even 
slightly better than without this state. It seems that this improvement of 
description can become more remarkable at an explicit consideration of
the $\eta2\pi$-threshold in the uniformizing variable, because the Breit--Wigner
analysis points to the considerable coupling of the $\rho$-like mesons with
the $\eta2\pi$ channel. However for the final conclusion the combined analysis of
several processes to which the investigated resonances are appreciably coupled have
to be performed.

{\it The fourth $\rho$-like meson turns out to have the mass about 1600 MeV rather
than 1720 MeV} cited in the PDG tables \cite{PDG-06}. Note that in a number 
of previous analyses of some reactions one has also obtained the resonance 
with mass near 1600 MeV \cite{PDG-06}. Note also a rather big obtained coupling 
of these $\rho$-like mesons with the 4$\pi$ channels.

As to the $\rho(1900)$, in this region there are practically no data
on the $P$-wave of $\pi\pi$ scattering. The model-independent analysis, maybe,
somehow testifies in favour of existence of this state, whereas the Breit--Wigner
one gives the same description with and without the $\rho(1900)$. For more definite
conclusion about this state, the $P$-wave $\pi\pi$ scattering data above 1880 MeV
are needed. Furthermore, the combined analysis of coupled processes should be 
carried out.

Finally, we note that the $P$-wave $\pi\pi$ scattering length ($a_1^1=33.9\pm
2.02~[10^{-3}m_{\pi^+}^{-3}]$), obtained in the Breit--Wigner analysis, matches most 
to the one in the local Nambu--Jona-Lasinio model \cite{BOM}.

\section{Acknowledgements}
Yu.S. acknowledges support provided by the Votruba-Blokhintsev Program for
Theoretical Physics of the Committee for Cooperation of the Czech Republic 
with JINR, Dubna. P.B. thanks the Grant Agency of the Czech Republic, 
Grant No.202/05/2142.

\end{document}